# Price Setting Rules, Rounding Tax, and Inattention Penalty*

Doron Sayag
Department of Economics, Bar-Ilan University
Ramat-Gan 5290002, Israel
Doronsayag2@gmail.com

Avichai Snir
Department of Economics, Bar-Ilan University
Ramat-Gan 5290002, Israel
Snirav@biu.ac.il

Daniel Levy**
Department of Economics, Bar-Ilan University
Ramat-Gan 5290002, Israel,
Department of Economics, Emory University
Atlanta, GA 30322, USA,
ICEA, ISET at TSU, and RCEA
Daniel.Levy@biu.ac.il

Revised:
November 19, 2024

* This is a substantially revised version of the manuscript we presented at the 6[th] Annual Conference on Law and Macroeconomics, held on November 2–3, 2024, at Tulane Law School, Tulane University. We thank the conference participants for their helpful and constructive comments and suggestions. The paper is based on Chapter 2 of Doron Sayag's PhD dissertation. All errors are ours.

Declarations of interest: None

** Corresponding author: Daniel Levy, Daniel.Levy@biu.ac.il

# Price Setting Rules, Rounding Tax, and Inattention Penalty

***Abstract***

We study Israel's "price rounding regulation" of January 1, 2014, which outlawed non-0-ending prices, forcing retailers to round 9-ending prices, which in many stores comprised 60%+ of all prices. The regulation's goals were to eliminate (1) the rounding tax—the extra amount consumers paid because of price rounding (which was necessitated by the abolition of low denomination coins), and (2) the inattention tax—the extra amount consumers paid the retailers because of their inattention to the prices' rightmost digits. Using 4 different datasets, we assess the government's success in achieving these goals, focusing on fast-moving consumer goods, a category of products strongly affected by the price rounding regulation. We focus on the response of the retailers to the price rounding regulation and find that although the government succeeded in eliminating the rounding tax, the bottom line is that shoppers end up paying more, not less, because of the regulation, underscoring, once again, Friedman's (1975) warning that policies should be judged by their results, not by their intentions.

"One of the great mistakes is to judge policies and programs by their intentions rather than their results."

**Milton Friedman**, *Living with Our Means*, Interview with R. Heffner on PBS's "Open Mind," Dec. 7, 1975[1]

## 1. Introduction

In 1991, the Bank of Israel (BOI) abolished the NIS 0.01 coin, leaving the NIS 0.05 coin as the smallest denomination. To enable cash transactions in which the final bill did not end in 5 or 0, the BOI adopted the "Swedish rounding" rule, requiring that in cash transactions, retailers round the final bill to the nearest NIS 0.05.[2] In 2008, the BOI abolished the NIS 0.05 coin as well. After that, the rounding rule was modified, requiring that the final bill be rounded to the nearest NIS 0.10. The modified rule drew criticism from consumer groups that felt that it was unfair, for its asymmetry: 4 endings (1–4) were rounded down, while 5 endings (5–9) were rounded up (Levy et al. 2011).

In October 2013, the government announced a new price rounding regulation: starting January 1, 2014, all prices must end with 0. I.e., since 2013, there are only ten possible price endings: .00, .10, … .80, .90. The Minister of the Economy defended the regulation, by stating that "For years, retailers profited twice – [1] they both presented a lower price [referring to 9-ending prices] that misled the consumers and, [2] collected excess amount from consumers [referring to the asymmetric price rounding rule]."[3]

In other words, the government argued that abolishing non-0-ending prices would eliminate two costs incurred by consumers. First, it would eliminate the "rounding tax"— the extra amount of money paid by consumers because prices are more likely to be rounded up than down (Lombra 2001). Second, it would stop retailers from using 9-ending prices to mislead shoppers into believing that prices are lower than they are (Schindler and Kibarian 1996, Anderson and Simester 2003, Thomas and Morwitz 2005).

In this paper, our goal is to assess whether the banning of non-0-ending prices has indeed achieved the two goals set by the government. Using data on fast-moving consumer goods (FMCG), we find that before the implementation of the price rounding regulations, the rounding tax was small, comprising about 0.001%–0.002% of the revenue in the FMCG market.

---

[1] The interview video recording is available at: https://www.thirteen.org/openmind-archive/public-affairs/living-within-our-means/. Friedman makes the above statement 1:54 / 28:38 into the interview, accessed on June 28, 2024.

[2] Similar rounding rules were adopted in other countries that abolished small denomination coins, such as Sweden, New Zealand, Australia, Finland, and the Netherlands (Leszkó 2009).

[3] Statement of Mr. Naftali Benett, who, at the time, was the Minister of the Economy. Source: https://www.kolhaemet.co.il/4174 (in Hebrew), accessed on November 28, 2023.

We also find that retailers responded to the elimination of 9-ending prices by setting a high share of 90-ending prices. Indeed, before the regulatory change that eliminated non-0 ending prices, the % of 9-ending prices was 63.2%, 19.9%, and 41.9% in supermarkets and drugstores, small grocery stores, and convenience stores, respectively. Once non-0-ending prices were outlawed, .90 became the dominant price ending: the % of 90-endings increased from 13.1%, to 71.8%, 57.0%, and 42.0%, in supermarkets and drugstores, small grocery stores, and convenience stores, respectively. I.e., after the banning of non-0-ending prices, the share of 90-ending prices became similar, or even higher, than the share that all 9-ending prices combined had before the price rounding regulation.

Previous research has shown that shoppers are often inattentive to the cent digits (Thomas and Morwitz 2005, Hodges and Chen 2022), and consequently, the demand elasticity of changes in the penny digits is low, giving retailers an incentive to set prices just below round numbers (Strulov-Shlain, 2023). After 99-ending prices were outlawed, the retailers increased the use of just-below prices in the form of 90-ending prices, which have de facto become the new just-below price.

Thus, while the government planned to eliminate the costs incurred by shoppers who were inattentive to the penny digits, we find that the high share of 90-ending prices after the elimination of non-0-ending prices led to an increase in the penny digits of the prices (e.g., an increase from NIS 2.19 to NIS 2.90). We estimate that by 2021, the increase in the number of pennies per price, cost shoppers about 0.27%–0.66% of the revenue in the FMCG market. Thus, of the two goals set by the government, it has only succeeded in abolishing the rounding tax. However, we find that the overall effect of the regulation was to increase the costs paid by shoppers, rather than decrease them.

The paper is organized as follows. In section 2, we offer a brief history of the NIS (the currency used in Israel) and the rounding regulations. In section 3, we review the literature on rounding tax, 9-ending prices, and the inattention penalty. In section 4, we describe the data. In section 5, we estimate the rounding tax. In section 6, we analyze changes in the ways stores use psychological price points and estimate the inattention penalty. In section 7, we conclude, offer caveats, and suggest avenues for future research.

## 2. Brief history of the New Israeli Shekel (NIS)

Prices in Israel are quoted in *New Israeli Shekels* (NIS).[4] Since 2000, and during our sample period, the average annual CPI inflation was 1.4%.[5]

The posted prices in Israel reflect the VAT and other taxes, and thus they are final prices. The currency denominations are 0.10, 0.50, 1, 2, 5, and 10 NIS coins, and 20, 50, 100, and 200 NIS notes. The NIS 0.01 (1-agora) and the NIS 0.05 (5-agora) coins were eliminated in 1991 and 2008, respectively, for their high cost of production.[6]

Although the NIS 0.01 and NIS 0.05 coins have not been in use since then, until 2014, retailers were free to use any price ending. If a consumer used a credit card, she paid the exact amount. If she paid with cash, however, the total bill was rounded. Until January 2008, prices were rounded to the nearest NIS 0.05. Between 2008 and 2014, prices were rounded to the nearest NIS 0.10. Therefore, in cash transactions between 1991 and 2008, a bill of NIS 9.42 was rounded down to NIS 9.40, a bill of NIS 9.45 remained unchanged, and a bill of NIS 9.48 was rounded up to NIS 9.50. Between 2008 and 2014, a bill of NIS 9.42 was rounded down to NIS 9.40, and bills of NIS 9.45 and NIS 9.48 were both rounded up to NIS 9.50.

The rounding rule adopted upon the elimination of the 5-agora coin drew criticism from consumer groups who argued that it was unfair because 5 price endings (5–9) were rounded up while only 4 price endings were rounded down (1–4). In other words, they argued the rounding rule was asymmetric, benefiting the sellers at the expense of buyers.

In addition, they argued that retailers set 9-ending prices, and especially 99-ending prices, to take advantage of shoppers who are inattentive to the rightmost digits. By setting 9-ending prices, they claimed the retailers were benefiting twice. First, shoppers perceive 9-ending prices as lower than the nearby round prices (e.g., NIS 4.99 is treated more like NIS 4.00 than NIS 5.00). Second, the rounding rule they used implied that in cash transactions, a 9-ending price was in fact the same as the nearby round price. E.g., a consumer that used cash to pay a price of NIS 4.99, paid NIS 5 and received no change.

---

[4] During our sample period, the average NIS/$ exchange rate was NIS 3.62 per $1.

[5] The standard deviation of inflation was 1.6%. For comparison, the average annual US (EU) CPI inflation in the same period was 2.2% (1.7%), with a standard deviation of 1.3% (1.2%).

[6] For example, by 2008 the cost of minting a 5-agora coin was 16-agora. Also, the public was reluctant to accept them as change. In addition, vending machines, parking meters, and other coin-operated devices stopped accepting them. See https://www.boi.org.il/en/information-and-service-to-the-public/my-cash-banknotes/the-bank-of-israel-asks-the-government-to-approve-abolishing-the-5-agorot-coin-as-legal-tender/, accessed on November 18, 2024.

A survey conducted by the Israel Consumers' Council in late 2013, illustrates the negative public sentiments towards 9-ending prices: (i) 63% of the survey participants perceived 9-ending prices as a ruse to make prices look lower, while (ii) 85% of the participants perceived the use of 9-ending prices as unfair.[7]

In response to public concerns, on October 19, 2013, the government announced that as of January 1, 2014, all non-0-ending prices were outlawed, leaving only 10 price endings that retailers could use: .00, .10, …, .80, and .90.[8] This regulation became known as the "price rounding regulation."

## 3. Literature review

As discussed above, the government's goals were to eliminate the rounding tax and to stop retailers from using 9-ending prices.

### 3.1. The rounding tax

Lombra (2001) argues that if small denomination coins are missing and final bills are rounded to the nearest available denomination, then shoppers will pay a penalty if the final bills are rounded up. He refers to this rounding penalty as the *rounding tax*.

The early literature on the rounding tax was motivated by an ongoing debate in the US on the merits of eliminating the 1¢ coin.[9] Lombra (2001) uses data from a US convenience store to estimate the "rounding tax" that shoppers would pay if the US abolished the 1¢ coin, and cash transactions would be rounded to the nearest 5¢. He assumes that convenience store shoppers buy 1–2 items, 82.5% of the prices end in 9, and the share of cash transactions is 50%–83%. Under these assumptions, he concludes that the rounding tax would amount to about $700 million/year.

Other papers find that the rounding tax is smaller, although not insignificant. Cheung (2018) finds that since the Royal Canadian Mint stopped producing the 1¢ coin in 2013, Canadian shoppers lost about CAD 3.3 million annually. This amount averages CAD 157

---

[7] Source: https://www.consumers.org.il/?catid={82C212BA-66D5-4EAC-9BE9-AA7A7F525675}, accessed on November 18, 2024.
[8] The prices of two product groups are exempted from the price rounding regulation. First, goods sold by weight, such as fresh fruits and vegetables, and fresh meat. However, the posted prices of these goods (price/kg) are usually set at 0-endings. Indeed, in our data (described below), in the period following January 1, 2014, 99.75% and 99.35% of the prices (per kg) of fresh fruits and vegetables, and fresh meat and fish, respectively, are 0-ending. The second group includes 21 narrowly defined (basic or essential) goods whose maximum prices are capped by the government. For these products, retailers are allowed to set the maximum prices, which are not necessarily round.
[9] See "Stop Making Cents," *New York Times – Sunday Magazine*, by Caity Weaver, Sept. 8, 2024, p. 26.

of additional revenue for a typical Canadian grocery store a year.

Keinsley (2013) uses a theoretical model and finds that if the 1¢ coin were eliminated, US shoppers would lose, on average, up to $1.6/year because of rounding. Thus, the total rounding tax paid by US shoppers would be up to $530 million/year.

Some studies report significantly smaller estimates of the rounding tax. Chande and Fisher (2003) review Lombra's (2001) result using price data from a Canadian fast-food restaurant chain. They assume that shoppers buy 1–4 items in their shopping trips, and add sales tax to the posted prices. Under these assumptions, they report that the rounding tax is positive but small and would "bear no consequence in everyday life" (p. 515).

Whaples (2007) uses data from a convenience store chain in 7 US states. Similar to Chande and Fisher (2003), he accounts for the local sales tax and concludes that in 6 of the 7 states, the rounding tax would benefit consumers rather than retailers. These predicted gains, however, are small, in the range of 0.01¢–0.10¢ per cash transaction.

Leszkó (2009) summarizes information on countries that have eliminated small denomination coins and concludes that the effect of eliminating small denomination coins is small.

### 3.2. 9-ending prices and the inattention penalty

It is believed that 9-ending prices (also known as "psychological price points" or "just-below prices") have a positive effect on sales (Kashyap 1995). For example, Anderson and Simester (2003) ran a controlled experiment with 3 mail-order catalogs, in which they manipulated the prices of an identical dress. They report that more orders were received through the catalog where the price was $39, than the orders received from the other two catalogs where the price was $34 and $44. Thus, people chose to buy more when the price was 9-ending, even though it was 14.7% higher than the lowest price. Other studies have also reported a positive effect of 9-ending prices on sales volumes (Stiving and Winer 1997, Schindler and Kirby 1997).

Studies suggest that the positive effect of 9-ending prices on sales volumes can be explained by shoppers perceiving 9-ending prices as lower than they are. This might be either because shoppers process price information left-to-right and pay little attention to the rightmost digits, which is known as the level effect (Schindler and Kirby 1997, Stiving and Winer 1997, Thomas and Morwitz 2005, Ruffle and Shtudiner 2006) or because they perceive 9-endings as a signal for low prices, which is known as the image effect (Schindler 1984, Stiving 2000, Schindler and Kibarian 1996, Quigley and Notarantonio 2015). For example, Levy, et al. (2020) find that in a lab experiment, participants who were asked to compare two prices were more likely to make a mistake when the rightmost digit of the higher price was 9.

In line with the view that 9-ending prices have a positive effect on sales, Levy et al. (2011), Knotek (2011), and Gorodnichenko and Talavera (2017) find that 9-ending prices are more rigid than prices ending with other digits (Blinder et al., 1999). Schindler (2001) and Snir and Levy (2021) report that the low-price image of 9-ending prices persists even though 9-ending prices are actually higher, on average, than non-9-ending prices.

Strulov-Shlain (2023) estimates a model of demand that assumes a left-to-right rounding bias (i.e., level-effect), and finds that shoppers have a low elasticity for within-dollar price changes. I.e., he finds that a 10¢ price increase from $5.89 to $5.99 has a smaller effect on demand than a 10¢ price increase from $5.99 to $6.09, or even a 1¢ increase from $5.99 to $6.00. He concludes that, on average, shoppers perceive 99-ending prices as 20¢ lower than the nearby round prices.

The effect of 9-ending prices on shoppers' behavior is not limited to the US. Ater and Gerlitz (2017), Snir et al. (2017), Levy et al. (2020), and Knotek et al. (2023) show that 9-ending prices were common in Israel before 2014, and offer evidence consistent with 9-ending prices having a positive effect on demand. Hoffmann and Kim-Kurz (2006) show that 9-ending prices are common and rigid in Germany. Gorodnichenko and Talavera (2017) use online data from the US and Canada and show that in both countries, 9-ending prices are more rigid than other prices, implying that retailers believe that 9-ending prices are more attractive than non-9-ending prices. Dhyne et al. (2006) report a similar finding for 10 EU countries, including Germany, France, Italy, and Spain. Gorodnichenko et al. (2024) show that 9-ending prices are common and rigid in Chinese online markets.

In sum, there is no consensus in the literature about the size of the rounding tax. Some studies find it to be a significant cost incurred by consumers, while other studies find it to be small and unimportant. There is a greater consensus about the costs associated with the prevalence of 9-ending prices. It seems that many shoppers perceive 9-ending prices as lower than they are, which means that they are paying higher prices than they perceive.

However, Strulov-Shlain (2023) argues that because 99-ending prices are particularly attractive, retailers might round some prices down to the nearest 99-endings. He, therefore, concludes that the effect of 99-endings on consumer welfare is ambiguous.

Other price endings can also become psychological pricing points. For example, Schindler (2009) and Gorodnichenko et al. (2024) report that in China and Japan, 8-ending prices are both common and rigid. Aalto-Setala (2005) shows that before Finland joined the Euro, 90 was the most common price ending in Finnish stores. Snir, et al. (2017) find that 6 months after the government passed the price rounding regulation, Israeli shoppers associated 90-endings with low prices. Strulov-Shlain (2021) also finds that Israeli shoppers perceive 90-ending prices as lower than the nearby round prices.

## 4. Data

We use four datasets. First, we use price data collected by the Israeli Central Bureau of Statistics (*CBS*) for compiling the CPI, during 2005–2021. The data includes price information for individual products offered in a sample of stores. The data is collected monthly by surveyors who visit the stores. We focus on Fast Moving Consumer Goods (*FMCG*), defined as products that are frequently purchased, quickly consumed, have low prices, and are sold in large quantities. We therefore focus on products sold in drugstores, supermarkets, small groceries, and convenience stores, that have a maximum price of NIS 200. This yields 4,341,643 monthly observations. The average price in our sample is NIS 28.85 with a standard deviation of 38.80. The median price is NIS 14.29.

Second, we use Nielsen's retail scanner data for 2011–2016, which contains monthly observations on revenue and sales volumes of products in the FMCG categories. The data are at the national level and are divided into three categories according to the type of retailer: supermarkets and drugstores, small groceries, and convenience stores.[10] The data is based on information obtained directly from individual retailers' POS scanners, which are then aggregated to generate the national-level data. Thus, an observation includes information on the total sales volume and revenue for a product (defined by UPC) in one of the three types of retailers.[11] Nielsen data covers about 96% of the FMCG market. The main group of outlets that is not represented in this data are stalls in open markets.

---

[10] In the original files, the data is divided to hypermarkets, supermarkets and drugstores, small groceries, convenience stores, and supermarkets targeting ultra-religious shoppers. To make it compatible with the CBS data, we pool the data on hypermarkets, supermarkets, and supermarkets targeting ultra-religious shoppers into one retailer category.

Third, we use information on consumers' shopping habits from the 2013 CBS household expenditure survey. The information comes from bi-weekly diaries that were kept by the 9,500 households that participated in the survey. In the diaries, households recorded every product purchased during a 14-day period, including information on the product, the store where it was purchased, and the price. In total, we have 870,000 observations on items bought. This dataset allows us to infer details such as the number of items purchased on each shopping trip.

Fourth, we use scanner data from a large national supermarket chain for 2011–2021. The data contains weekly observations on revenues and sales volumes of 21,831 food and household products, where a product is defined by its UPC. The average price in this dataset is NIS 17.61 with a standard deviation of NIS 10.85. The average sales volume per store, per week, per UPC, is 5.76 units, with a standard deviation of 9.25. Below, we refer to this dataset as the *retailer's* dataset.

## 5. The rounding tax

To calculate the rounding tax, i.e., the extra amount consumers pay because prices are rounded up, we follow Lombra (2001), Chande and Fisher (2003), and Whaples (2007). However, our datasets include information on the distribution of price endings and the size of shopping baskets, information that was not available to the above authors. We can, therefore, offer a more precise estimate of the rounding tax. In addition, our data allows us to measure the rounding tax separately for different types of stores.

To start, we use the CPI data to calculate the shares of price endings by type of stores: supermarkets and drugstores, small grocery stores, and convenience stores.[12] We compute the shares for 2013, as this is the final year before non-0-ending prices were abolished.

We focus on FMCG products because these products have relatively low prices and are bought in large numbers and with high frequency. Therefore, these are the products that are most likely to be relevant for the rounding tax. Indeed, Levy, et al. (2011) show that as prices get higher, the 9-digits tend to shift to the left, implying that the right-most digits of relatively high prices are usually zero.

---

[11] For illustration, an observation may state that in a given month, a particular brand of orange juice, identified by its UPC, sold in convenience stores 19,543 units, for a total revenue of NIS 140,123.30.

[12] We do not separate the information on supermarkets and drugstores because Nielsen treats them as a single group.

Second, we use information from the 2013 household expenditure survey to calculate the distribution of basket sizes for each type of store. Figure 1 depicts the distributions of the price endings, and of the size of shopping baskets for each of the three types of stores.

There is large heterogeneity across store types in the distributions of both the price endings and the shopping baskets. Consistent with Knotek (2011) and Snir et al. (2022), 9 is the most common price ending in supermarkets and drugstores, while 0 is the most common price ending in small grocery stores and convenience stores. In supermarkets and drugstores, 9-ending and 0-ending prices comprise 61.1% and 18.3% of the prices, respectively. In contrast, in small grocery stores, 9-ending and 0-ending prices comprise 19.1% and 75.8% of the prices, respectively, while in convenience stores, they comprise 34.2% and 64.1% of the prices, respectively.

Looking at basket sizes, we find that in convenience stores, 71.0% of the transactions involve exactly 1 good. This is similar to the figures reported by Lombra (2001) for the US and Chande and Fisher (2003) for Canada. In supermarkets and drugstores, 13.6% of the shopping baskets are size one, the median basket size is 6, while 25% of the shopping baskets contain 15 or more products. Small grocery stores are in between, with the median shopping baskets containing 3 products.

To estimate the rounding tax, we use the distributions of the price endings along with the distributions of the basket sizes to simulate 10,000 transactions for each of the three types of retailers (Lombra 2001, Chande and Fisher 2003). In each simulation, we employ a two-stage procedure. In the first stage, we determine the size of the shopping basket according to the likelihood of observing each basket size. In the second stage, we determine the price ending of each product in the basket according to the distribution of price endings in that type of store. We then calculate the rounding tax for the transaction and find the average over 10,000 transactions.

Column 1 of Table 1 gives the simulation results for the average rounding tax per cash transaction for each of the three store types. The average rounding tax is the highest in drugstores and supermarkets, NIS 0.0075 per transaction. The lowest average rounding tax is found in convenience stores, NIS 0.0048 per transaction. In small grocery stores, the average rounding tax is NIS 0.0058 per transaction.

Column 2 gives the share of each type of store in the total FMCG revenue in 2013, based on Nielsen's data. Column 3 gives the total number of FMCG transactions in 2013 per type of store, which we calculate by taking the total number of transactions for each of the three types of stores from Nielsen's dataset and dividing it by the average number of items per shopping trip at the corresponding type of store.

If shoppers were to pay the rounding tax for all transactions, then we could calculate the total rounding tax per type of store by multiplying columns (1) and (2). However, shoppers pay the rounding tax only for cash transactions. To account for this, we combine data on total expenditure and on payments by credit card in the FMCG market. Using the combined data, we estimate that the share of cash transactions out of all FMCG transactions in 2013 was 25%.[13]

Since we do not have data about the utilization of credit cards by store type, Column 4 reports our baseline estimate of the total rounding tax for 2013 assuming that the share of cash transactions was 25% in all stores. Under this assumption, we find that the total rounding tax paid by Israeli shoppers in 2013 was NIS 507,280.

However, it is unlikely that the share of cash transactions in supermarkets was the same as in convenience stores. Indeed, Bouhdaoui et al. (2014), Chen et al. (2019), and Shy (2020) show that the share of cash transactions is higher in convenience stores than in supermarkets and drugstores. We therefore perform the following calculation. We allow the share of cash transactions to vary across the types of stores, such that the overall share of cash transactions remains 25%. We then find the shares of cash transactions that maximize (minimize) the overall rounding tax.

We find that the maximum (minimum) rounding tax is obtained when the share of cash transactions in small grocery stores and convenience stores is 100% (0%), implying that the share of cash transactions in supermarkets and drugstores is 10.6% (29.8%). The total rounding tax paid in 2013 in the maximum (minimum) scenario is NIS 763,641 (NIS 422,390).

---

[13] To estimate the share of cash transactions, we use two data sources. First, we have total expenditures in the FMCG market from the Nielsen's dataset. Second, we use information based on data collected by the CBS on total credit card expenditures in the FMCG market. The difference between total expenditures and total credit card expenditures is our estimate of the share of cash transactions in the FMCG market.

According to Nielsen's data, the total revenue in the FMCG market in 2013 was NIS 40.8 billion. Our results, therefore, suggest that in that year, the rounding tax was 0.001%–0.002% of the total revenue.

## 6. Psychological price points and inattention penalty

Strulov-Shlain (2021, 2023) suggests a model that can capture the left-digit bias associated with 9-ending prices using a simple framework. He assumes that shoppers process price information left-to-right and pay less attention to the right-most digits, such that a product's perceived price, $\hat{p}$, is given by:

$$\hat{p} = (1 - \theta)p + \theta(\Delta + \lfloor p \rfloor) \tag{1}$$

where $\lfloor \cdot \rfloor$ is the floor operator, $p$ is the product's true price, $\Delta \in [0, 0.99]$ is the focal price ending, and $\theta \in [0, 1]$ is the left-digit bias parameter. As $\theta$ gets larger, shoppers pay less attention to the true price and more attention to the left-most digit. For example, with $\Delta = 0$ and $\theta = 1$, a price of 9.99, for example, is perceived as $(1 - 1)9.99 + 1(0 + 9.00) = 9.00$. With $\theta = 0.2$, A price of 9.99 is perceived as $(1 - 0.2)9.99 + 0.2(0 + 9.00) = 9.79$. Thus, when the left digit bias, $\theta$, is 0.2, a shopper underestimates a price of NIS 9.99 by NIS 0.20. When the left digit bias, $\theta$, is 1, a shopper underestimates a price of NIS 9.99 by NIS 0.99.

Strulov-Shlain (2023) finds that in the US, the bias parameter is about 0.2. As discussed above, Israeli shoppers perceived 99-ending prices as "unfair" and as a ruse to make prices seem lower than they are. This suggests that they, like US shoppers, have a left-digit bias. Consistent with this hypothesis, Strulov-Shlain (2021) uses data on a small sample of products offered in Israeli supermarkets and finds that the left-digit bias of Israeli shoppers is also around 0.2.

Our goal is to study the effect of outlawing 9-ending prices on shoppers' welfare. As a first step, therefore, we test whether we can confirm Strulov-Shlain's findings. To do so, we use the retailer's dataset to estimate the left-digit bias.

Following Strulov-Shlain (2021), we take advantage of the discontinuity resulting from the price rounding regulation to estimate the left-digit bias. Consider a product whose price in 2013 was NIS 9.99, and on January 1, 2014 the price was either increased by NIS 0.01 (to NIS 10.00) or decreased by NIS 0.09 (to NIS 9.90), because the store had to comply with the price rounding regulation. Any change in the demand between these two prices (i.e., between NIS 9.90 and NIS 10.00) that is above and beyond the change implied by the price elasticity, can be attributed to a left-digit bias. We therefore estimate:

$$Ln(Q_{i,s,t}) = \alpha + \beta^{90}D_{i,s,t}^{90} + \beta^{00}D_{i,s,t}^{00} + \beta^{99}D_{i,s,t}^{99} + \varepsilon ln(P_{i,s,t}) + X'\gamma + u_{i,s,t} \qquad (2)$$

where $Q_{i,s,t}$ is the quantity of product $i$ sold in store $s$ in week $t$, $P_{i,s,t}$ is the price of product $i$ sold in store $s$ in week $t$, and $D_{i,s,t}^{90}, D_{i,s,t}^{00}$ and $D_{i,s,t}^{99}$ are dummy variables defined as follows. First, we focus on 2013, and for each product-store pair, we find the modal 99-ending price, $Mode2013_{i,s}^{99}$. We then set $D_{i,s,t}^{99}$ equal to 1 if in 2013 $P_{i,s,t}$ was equal to $Mode2013_{i,s}^{99}$, and 0 otherwise. We set $D_{i,s,t}^{00}$ equal to 1 if in 2014 $P_{i,s,t}$ was NIS 0.01 higher than $Mode2013_{i,s}^{99}$, and 0 otherwise. We set $D_{i,s,t}^{90}$ equal to 1 if in 2014 $P_{i,s,t}$ was NIS 0.09 lower than $Mode2013_{i,s}^{99}$, and 0 otherwise. For example, if the modal 99-ending price of a product-store in 2013 was NIS 9.99, then $D_{i,s,t}^{99}$ equals 1 in each week of 2013 in which the price was NIS 9.99, $D_{i,s,t}^{00}$ equals 1 in each week of 2014 in which the price was NIS 10.00, and $D_{i,s,t}^{90}$ equals 1 in each week of 2014 in which the price was NIS 9.90. The vector of control variables $X$ includes fixed effects for product-stores, for product categories × years, for product categories × months, and for supermarket chains.

The retailer dataset is not ideal for this type of estimation, because it is scanner data, which means that the prices we have are not the posted prices, but weekly averages, calculated as total revenue over total sales (Eichenbaum et al., 2014). To overcome this limitation, we exclude observations if the average price has more than 2 digits after the decimal point.[14] To demonstrate robustness, in the appendix, we present results where we include only observations where we also require that the price remained unchanged for at least two consecutive weeks. Strulov-Shlain (2023) shows that it is very unlikely that a spurious price would remain unchanged for more than 1 week.

Strulov-Shlain (2023) limits his estimation to prices below US$12, arguing that higher prices are "big-ticket items." For our main sample, we use prices up to NIS 20. Products in this price range comprise 72% of the prices in the retailer dataset and are

---

[14] After 2014, we exclude observations also if the two rightmost digits are not a multiple of NIS 0.10.

responsible for 88% of the revenues.[15]

Table 2 contains the estimation results of equation (1). Column 1 reports the results when we include only products that satisfy the following three criteria: (1) their price was set at a 99-ending in 2013, (2) in 2014, at least once, their price was NIS 0.01 higher than the 2013 modal 99-ending price, and (3) in 2014, at least once, their price was NIS 0.09 lower than the 2013 modal 99-ending price. Thus, we restrict the sample to product-store pairs, where the price endings fluctuated in 2013 and 2014 between 99, 90, and 00.

We find that as predicted, the coefficient of $D_{i,s,t}^{90}$ is higher than the coefficient of $D_{i,s,t}^{00}$, suggesting that a 90-ending price leads to higher demand than the nearby 00-ending price, and the difference is larger than predicted by the demand elasticity. We follow Strulov-Shlain (2021) to estimate the left-digit bias parameter $\hat{\theta} \approx \frac{(\hat{\beta}^{90} - \hat{\beta}^{00})}{-\hat{\varepsilon}} \bar{p}$, where $\bar{p}$ is the average price, and find that it equals 0.22, very close to what Strulov-Shlain (2023) reports for the US.

In column 2, we restrict the sample to products with prices less than NIS 10, because for them the left-digit bias is likely to be the strongest. We find that the left-digit bias parameter equals 0.30, which is indeed higher than what we obtained for the sample where the maximum price was NIS 20.

In column 3, we again restrict the sample to products with prices less than NIS 20. We also eliminate restrictions (2) and (3), and instead include only products whose price in 2014 was at least once NIS 0.01 higher than the 2013 modal 99-ending price, or whose price in 2014 was at least once NIS 0.09 lower than the 2013 modal 99-ending price. This increases the sample considerably, from 994,459 observations to 6,947,640 observations. However, this estimation introduces noise, because we compare the effect of price changes across products rather than within products. In this sample, we find that $\hat{\theta} \approx$ 0.69, which seems too high.

In column 4, we repeat the specification of column 1 but stop the sample period in December 2019, because the 2020 data might have been affected by the COVID-19 lockdowns. With this restriction, we get $\hat{\theta} \approx 0.11$, lower than we obtained in the full

---

[15] This is a less restrictive sample than Strulov-Shlain (2021) uses, because the average price in his sample is NIS 11, compared to NIS 12.70 in our sample.

sample, but still considerable.

Thus, we conclude that there is a left-digit bias in Israel and its size is similar to the US. Strulov-Shlain (2021, 2023) argues that with a bias of the magnitude we obtain, all prices in Israel before January 2014 should have been 99-ending and all prices after January 2014 should have been 90-ending.

However, retailers likely have constraints on the share of 99/90 ending prices they can use. For example, convenience stores' managers are likely to set many round prices to accommodate their shoppers' preference for convenience (Schindler and Kirby 1997, Knotek 2008 and 2011). Similar considerations may apply to prices of products offered in drugstores and some supermarket departments (Snir et al., 2022), where they might prefer round prices for some product categories such as front-end candies, or they might want to convey a high-quality image (Stiving 2000, Mastrobuoni et al. 2014). We, therefore, expect that the share of 99-ending prices would be less than 100%.

Figure 2 depicts the shares of prices with endings in the range 90–99 in 2005–2021 in the three types of stores. The vertical line marks January 1, 2014, the date on which the "price rounding regulation" came into effect. The share of 90–99 ending prices is high throughout the sample period in all three types of stores. In 2005–2013, the share of 90–99 ending prices in supermarkets and drugstores was stable at 50.6%–56.5%. In small grocery stores, the share increased between 2005 and 2010 and then stabilized at 43.7%–46.5% in 2011–2013. In convenience stores, the share rose between 2005 and 2010 from 20.5% to 29.5%, jumped to 42.2% in 2011, and then decreased to 25.4% in 2013.

In January 2014, following the price rounding regulation, all endings in the range of 90–99 were converted to a single ending, 90. This resulted in a dip in the share of 90–99 ending prices in supermarkets and drugstores. There was also a small dip in the share of 90–99 ending prices in small grocery stores. However, following 2015, the share of 90-ending prices started to increase in supermarkets and drugstores, and also in small grocery stores. In convenience stores, the share of 90-ending prices started to increase in 2017. By 2021, the share of 90-ending prices in supermarkets and in drugstores (small grocery shops) was 15.3 (10.6) percentage points higher than the highest share of 90–99 ending prices before 2014. In other words, in supermarkets and drugstores, the share of 90-ending prices in 2021 was 27% higher than the pre-2014 share of all 90–99 ending prices together. In small grocery stores, the share of 90-ending prices in 2021 was 23% higher than the corresponding pre-2014 share.

In convenience stores, the share of 90-ending prices in 2021 was 42%, about the same as in 2011. If we ignore 2011–2012 which had a particularly high share of 90–99 ending prices, then the share of 90-ending prices in 2021 is 12.5 percentage points (42%) higher than the highest share of 90–99 ending prices before 2014.

These results suggest that retailers responded to the elimination of 99-ending prices by increasing the share of 90-ending prices. Further, before 2014 the share of 90–99 ending prices seemed to have reached a plateau, at least in supermarkets and drugstores, as well as in small grocery stores. In 2014–2021, on the other hand, the share of 90-ending prices was continuously increasing.[16]

One possible explanation of this finding is that retailers, who know that shoppers have a left-digit bias, compensate themselves for the 9-agora loss that results from adopting 90-ending prices to replace 99-ending prices, by increasing the share of 90-ending prices. This way, they earn NIS 0.09 less, but now they make more 90-ending price transactions, than 99-ending price transactions they made before the change.

Another possibility is that the price rounding regulation cut the number of pricing options available. Before 2014, retailers had 100 possible price endings 00–99, which included ten 9-ending price options: 09, 19, …, 89, and 99. After 2014, retailers that wanted to keep using "9-ending prices," had only one option, 90. In other words, a retailer that wanted to keep using 9-ending prices was forced to adopt 90-ending prices. Thus, 90-ending is a substitute not for 99-ending prices only, but rather for all 9-ending prices.

This hypothesis seems to be consistent with the data. The share of 9-ending prices before 2014 was 63.2%, 19.9%, and 41.9% in supermarkets and drugstores, small grocery stores, and convenience stores, respectively. In 2021, the share of 90-ending prices was 71.8%. 57.0%, and 42.0% in supermarkets and drugstores, small grocery stores, and convenience stores, respectively. Therefore, in supermarkets and drugstores as well as in convenience stores, the share of 90-ending prices in 2021 is comparable to the share of all 9-ending prices prior to 2021. Small grocery stores seem to use 90-ending prices even more than they used 9-ending prices.

---

[16] The finding that the share of 90-ending prices was still increasing in 2021, 7 years after the price rounding regulation came into effect is consistent with Strulov-Shlain's (2021) argument that retailers learn on the run (Cross 1973, Rao and Simonov 2021).

We should note a caveat, however. Retailers most likely round prices both up and down to 90-ending prices. For example, an increase in the share of NIS 9.90 prices might be driven both by prices in the region NIS 9.00–9.80 being rounded up to NIS 9.90, or by prices in the region NIS 10.00–10.80 being rounded down to NIS 9.90.

If the increase in the share of 90-ending prices is driven by more prices being rounded down, then the effect on shoppers' welfare is ambiguous. That is because on the one hand, rounding prices down increases shoppers' welfare because they pay lower prices. On the other hand, however, because shoppers have a left-digit bias, they buy more than their demand function dictates, which lowers their welfare.

However, there are theoretical and empirical reasons to believe that the change in the share of 90-ending prices following 2014 was mostly driven by more prices being rounded up than by more prices being rounded down. First, from a theoretical point of view, retailers lose more when they round down to a 90-ending than to a 99-ending. For example, prior to 2014 a retailer might have preferred setting a price of NIS 9.99 over NIS 10.10 because the higher demand at NIS 9.99 compensated for the loss of NIS 0.11 per transaction. Following 2014, however, it is not clear that the higher demand at NIS 9.90 would compensate for losing NIS 0.20 per transaction. If retailers were to mostly round prices down, therefore, the share of 90-ending prices after 2014 should have been lower than the share of 99-ending prices before 2014. We find, however, that the share of 90-ending prices after 2014 is actually higher than the share of all 9-ending prices before 2014, not only higher than the share of 99-ending prices.

Second, empirically, Figure 3 depicts the shares of 2-digit endings. To allow comparisons between the distributions before and after 2014, we divide the price-endings into 10-agora segments: 00–09, 10–19, …, and 90–99. The LHS panel gives the distributions of the price endings in 2013, a year before the change in the regulation, and in 2021, the last year for which we have data. The RHS panel gives the percentage change in the share of each segment between 2013 and 2021.[17] If prices were rounded down to 90-endings, we should have mainly seen a decrease in the shares of price-endings just above the 90-ending group. In particular, we should have seen a decrease in the share of 00–09 ending prices. However, we find that in small grocery stores and convenience stores, the shares of 00–09 ending prices have hardly changed between 2013 and 2021. In supermarkets and drugstores, the share of 00–09 ending prices has decreased, but the relative decrease is much smaller than the relative decrease in the shares of other segments. Therefore, it seems that whereas rounding down to 90-ending prices after 2014 should have resulted in "missing price endings" (Strulov-Shlain, 2023) above 90-endings, this did not happen in small grocery stores and convenience stores. It has happened to a limited degree in supermarkets and drugstores. It, therefore, seems that the main effect of eliminating 99 and other non-0-ending prices was to increase the share of prices that were rounded up to 90-ending prices, rather than to increase the share of prices that were rounded down to 90-ending prices.

In the appendix, we present more calculations that suggest that the increase in the share of 90-ending prices is indeed driven by "rounding" prices up to a 90-ending rather than by rounding prices down. If we assume that retailers set prices at 90-endings when they forgo larger price increases in favor of keeping prices that maintain the same LHS (i.e., full NIS) digit(s), then the size of price increases to 90-ending prices should be smaller, on average, than the size of price increases to other endings. However, we find that on average, when a price increases to a 90-ending price, the average increase is larger than when a price increase is to a non-90-ending.

Thus, the results of the robustness test support the hypothesis that the increase in the share of 90-ending prices is an outcome of an increase in the penny digits of the price, yielding a higher average number of pennies per price. I.e., following the price rounding regulations, retailers that decide to set a price with a given left-most digit became more likely to set 90 as the rightmost ending rather than some other, lower price ending.

We, therefore, conclude that following the implementation of the price rounding regulation in 2014, retailers have increased the share of 90-ending prices, leading to an increase in the average number of pennies per price. Figure 4 shows the effect. We focus on prices up to NIS 20, as these prices are most likely to be affected by the left-digit bias, and they comprise the bulk of the prices in the FMCG market. In the appendix, we present the results for prices greater than NIS 20.

---

[17] It is possible that the distribution of price endings in 2013 might have been affected by public discussions about the price rounding regulation. In the appendix, we replicate the analysis using data for 2012. The results are similar.

In all types of stores, before 2014, the maximum number of agora (pennies) the shoppers paid per price was reached in 2011–2012 and then decreased in 2013 and 2014. Following 2014, there was a gradual and continuous increase in the average number of agora (pennies) the shoppers paid per price. By 2021, in supermarkets and drugstores and in small grocery stores, the average number of agora (pennies) that a shopper pays per transaction is higher than she paid before 2014.

We can calculate the effect of this on the total sum paid by shoppers, by computing the difference in the average number of agora (pennies) paid between the period before and after the price rounding regulation, multiplied by the number of products bought. Table 3 gives the estimation results, using information on prices up to NIS 20.[18] In Table 3, we compare the average prices in 2021 with those of 2013. Because the data for 2013 might have been affected by the anticipation of the government's plan to announce the price rounding regulation, in the appendix we present the results for 2012.

The first two rows give the average number of agora (pennies) per price in 2021 and 2013. The third row gives the differences between the two years. We find an increase of between 3.3–8.6 pennies in the number of pennies per price , depending on the type of store. The fourth row gives the sales volumes for each type of store, which we take from the 2013 Nielsen data. The final row gives the total effect on the amount paid by shoppers, which we obtain by multiplying rows 3 and 4.

The results reported in Table 3 suggest that the increase in the average number of pennies per price has a significant effect on shoppers' expenditures. The total additional amount paid by shoppers because of the increase in the number of agora (pennies) paid per price is NIS 269,185,684, which is about 0.66% of the total revenue in the FMCG market. If we use the data for 2012 (see the appendix) then the total effect is smaller, but still considerable, NIS 110,733,171, or about 0.27% of the total revenue.

---

[18] In the appendix, we present the results for prices that exceed NIS 20.

### 7. Conclusions and Caveats

We use data from Israel, a country where the NIS 0.01 and the NIS 0.05 were abolished in 1991 and 2008, respectively. Until January 2014, the retailers were allowed to use any price ending. In cash transactions, the final bills were rounded to the nearest available denomination, as in Sweden, Canada, Australia, and other countries that abolished small denomination coins (Leszkó, 2009).

On January 1, 2014, the government enacted a "price rounding regulation" outlawing all non-0-ending prices. By enacting this regulation, the government hoped to (1) eliminate the rounding tax paid by shoppers, and (2) prevent retailers from using 99-ending prices, which consumers viewed as manipulative and unfair.

We estimate that before the change in regulations, the value of the rounding tax was 0.001%–0.002% of the revenue in the FMCG market, which seems small, particularly when considering that the BOI abolished both the NIS 0.01 and the NIS 0.05 coins. In Israel, therefore, 5 price endings (5–9) were rounded up but only 4 price endings (1–4) were rounded down. Thus, the rounding tax in Israel is likely to be higher than in countries that abolished only the 1 penny coin, leading to 4 price endings that are rounded up and 4 price endings that are rounded down.

In addition, when the price rounding regulation came into effect in January 2014, about 25% of all transactions in the FMCG market were paid with cash. The advances in electronic payment methods in the period since 2014 likely resulted in a decrease in the share of cash transactions, implying that the rounding tax is becoming smaller over time.

As for the elimination of 99-ending prices, we find that although the price rounding regulation ended the retailers' ability to use non-0-ending prices, this has not necessarily eliminated "9-endings." It seems that although the government banned 9-ending prices, the retailers wanted to maintain the aura that 9-endings bestow on prices. However, whereas before the regulation the retailers had 10 possible 9-ending prices: 09, 19, …, 89, and 99, after the regulation came into effect, the retailers had only one alternative, 90. The end result is that the share of 90-ending prices became similar, or even higher than the share of all 9-ending prices before the enactment of the price rounding regulation. Thus, the government's attempt to eliminate "just below" prices has backfired. Instead of eliminating these prices, the price rounding regulation made just below prices even more popular than they were prior to the regulatory change.

In addition to maintaining the aura of "9-endings," the increase in the share of 90-ending prices has benefited the retailers because it led to an increase in the average number of pennies per price. Thus, for a price with the same LHS (i.e. full NIS) digits, after the price rounding regulation, the shoppers pay on average a few pennies more than they paid before.

Thus, instead of the new regulation discouraging retailers from taking advantage of the shoppers' left-digit bias (Thomas and Morwitz, 2005), it seems that it has encouraged them to take advantage of it to a greater extent than before. This result is consistent with Snir et al. (2017), who find that after January 2014 shoppers learned to associate 90-ending prices with a low-price image, similar to the association between 9-ending prices and a low-price image that existed before 2014. It is also consistent with Strulov-Shlain (2021), who argues that Israeli retailers should take advantage of the shoppers' left-digit bias by setting all prices at 90-endings.

The outcome is an extra cost incurred by shoppers. We estimate this cost is 0.27%–0.66% of the revenue in the FMCG market, much higher than the cost that the shoppers saved from eliminating the rounding tax. Even if we assume that this estimate is on the higher side because some of the increase in the share of 90-ending prices might be due to prices that were rounded down rather than up, the effect of the rounding up would still be much larger than the effect of eliminating the rounding tax.

There are three reasons for this finding. First, the increase in the average number of agora (pennies) per price affects the prices of all products whereas the rounding tax is calculated for the final bill only. Second, the higher average agora (pennies) per price is paid whether a shopper uses cash or an electronic payment method, whereas the rounding tax is paid only for cash transactions. Third, the average size of the rounding tax is less than 1 agora (penny) per final bill, whereas the increase in the average number of agora (pennies) paid per price is several agora per product bought.

We conclude that retailers' reaction to the price rounding regulation made the overall effect of the regulation on shoppers' welfare negative. Following the implementation of the price rounding regulation, the shoppers pay more, not less. In addition, the elimination of non-0-ending prices increases the distance between possible prices, making small price changes less likely. This can affect the likelihood of price changes, which might exacerbate market inefficiencies. Sandler et al. (2024) report findings that are consistent with this hypothesis. They find that in the period 2018–2021, regular price changes in Israeli supermarkets were rare, and small price changes were even rarer.

The elimination of 9-ending prices might have had an additional negative effect on shoppers' welfare. Many macroeconomic models assume that changing a price entails a fixed "menu cost" (Mankiw 1985, Levy et al. 1997 and 1998, Dutta et al. 1999, Zbaracki, et al. 2004). This suggests that retailers might have perceived the passing of the price rounding regulation in January 2014, which forced them to change a large share of the prices in their stores, as costly. It is possible that retailers tried to recoup some of the costs by either increasing some prices or by foregoing price decreases. Future research can shed more light on this issue.

Thus, our results suggest that policies should indeed be judged by their effects rather than their intentions, to paraphrase Friedman's (1975) opening quote. The "rounding regulation" was enacted in response to consumer groups demands which called for banning 9-endings prices. The final result, however, is that consumers pay more, not less.

We should note the following caveat, however. If shoppers strongly oppose the use of 9-ending prices, as suggested by surveys conducted before the government enacted the price rounding regulation, then the elimination of 9-ending prices might have still made the shoppers better off even if they pay, on average, a few pennies more per transaction. Future research may study the welfare effects of the elimination of 9-ending prices by assessing the price that shoppers are willing to pay for eliminating the use of 9-ending prices. Then we could find whether the shoppers' willingness to pay is indeed higher or lower than the cost that they eventually ended up paying.

Figure 1. The distributions of price endings and basket sizes in 2013

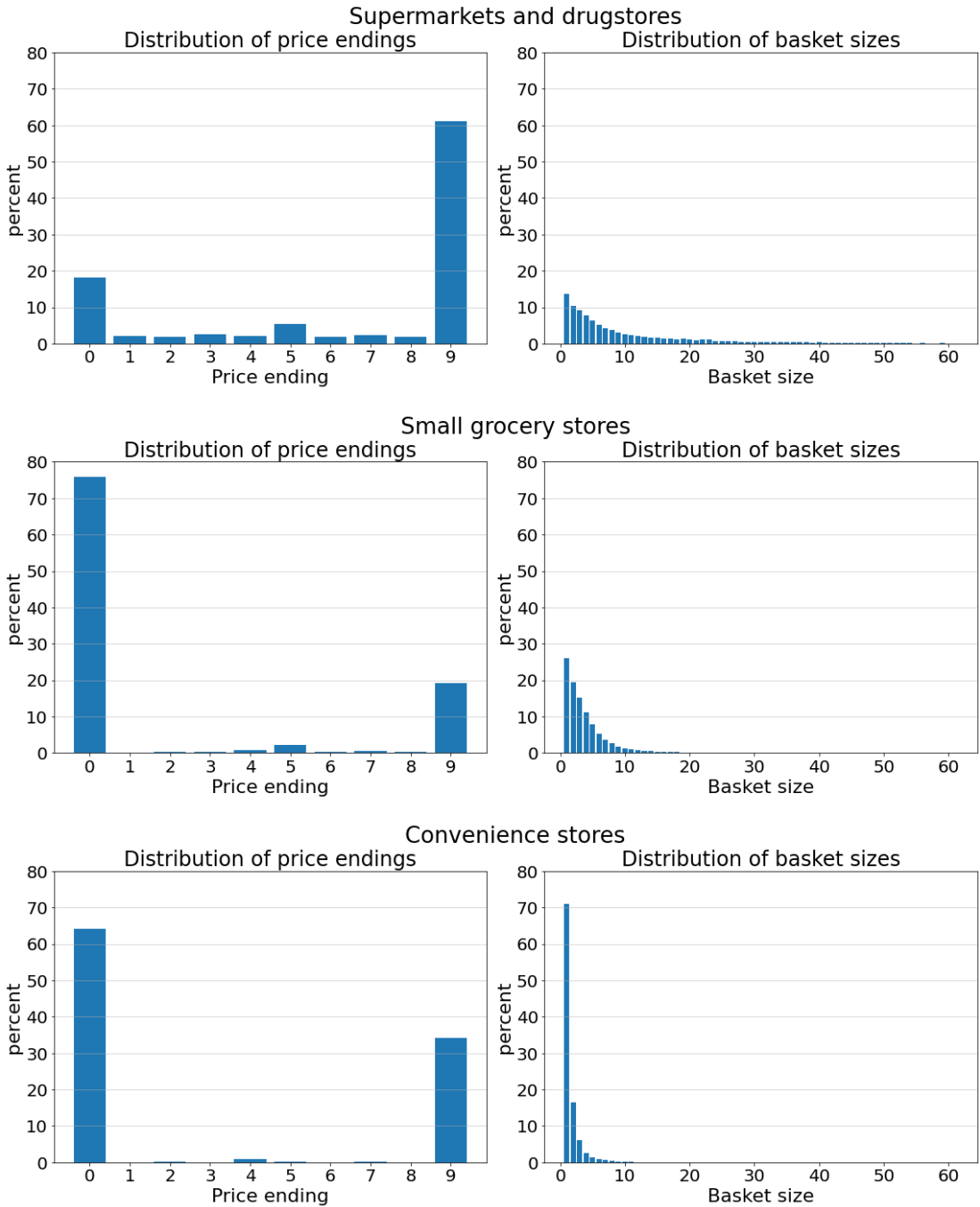

Figure 2. The shares of 90–99 ending prices

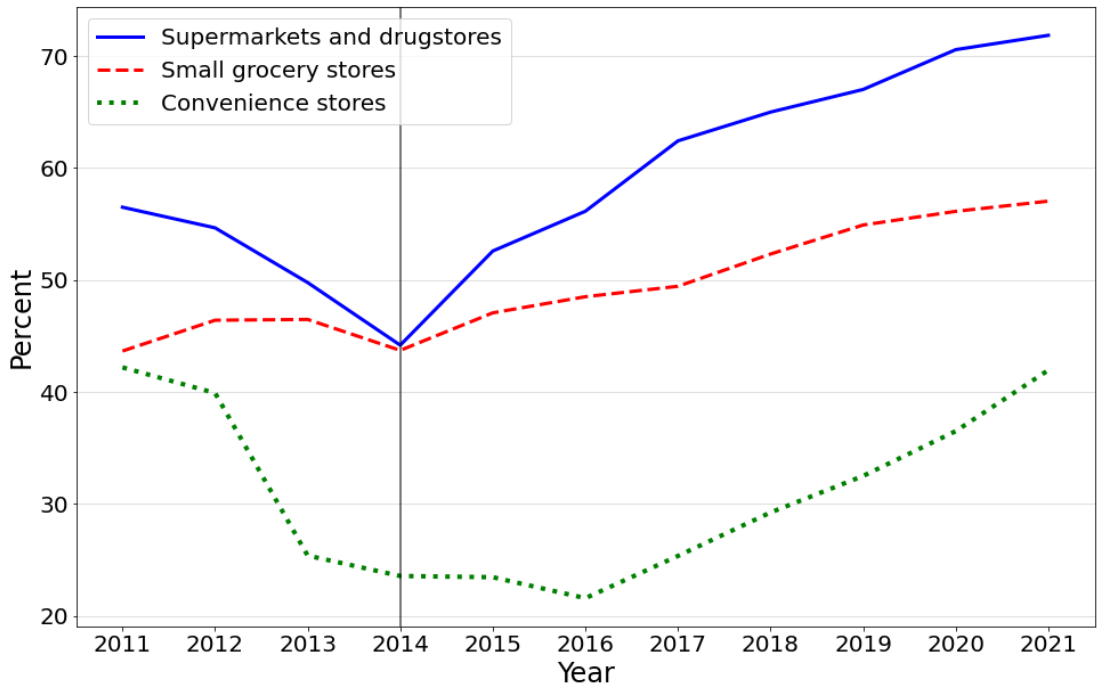

Notes: The shares of 90–99 ending prices in supermarkets and drugstores, small grocery stores, and convenience stores. Data based on CPI sample. The vertical line marks January 1, 2014, when the price rounding regulation came into effect.

Figure 3. The distribution of the last two digits, 2013 vs. 2021

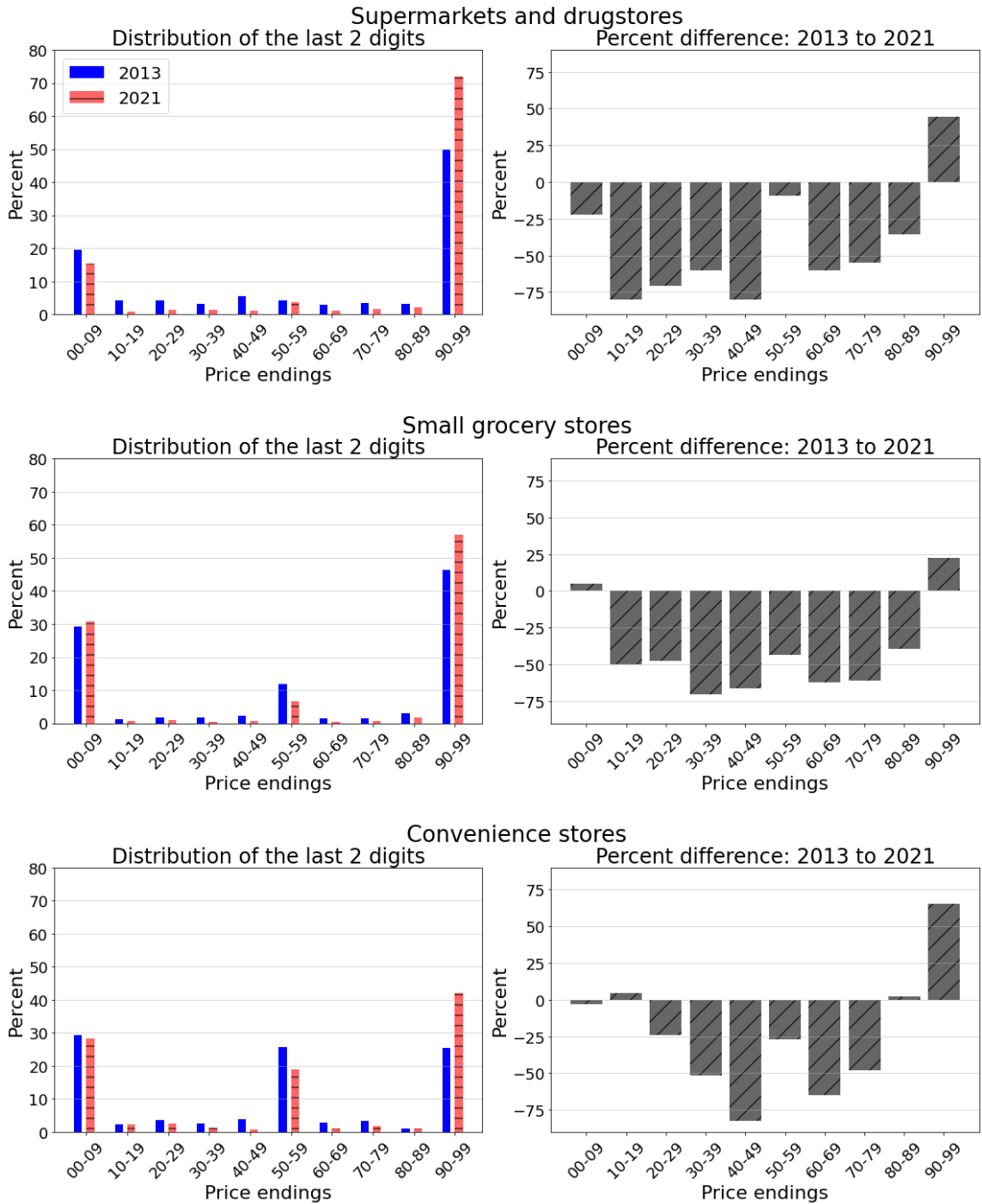

: The LHS panel depicts the distribution of the last two digits in 2013 and 2021. To allow comparison, we bunch price endings into 10-agora segments: 00–09, 10–19, …, and 90–99. The RHS panel depicts the percentage change in the shares of each segment between 2013 and 2021. The Data is based on the CPI sample.

Figure 4. Average number of agora (pennies) per price

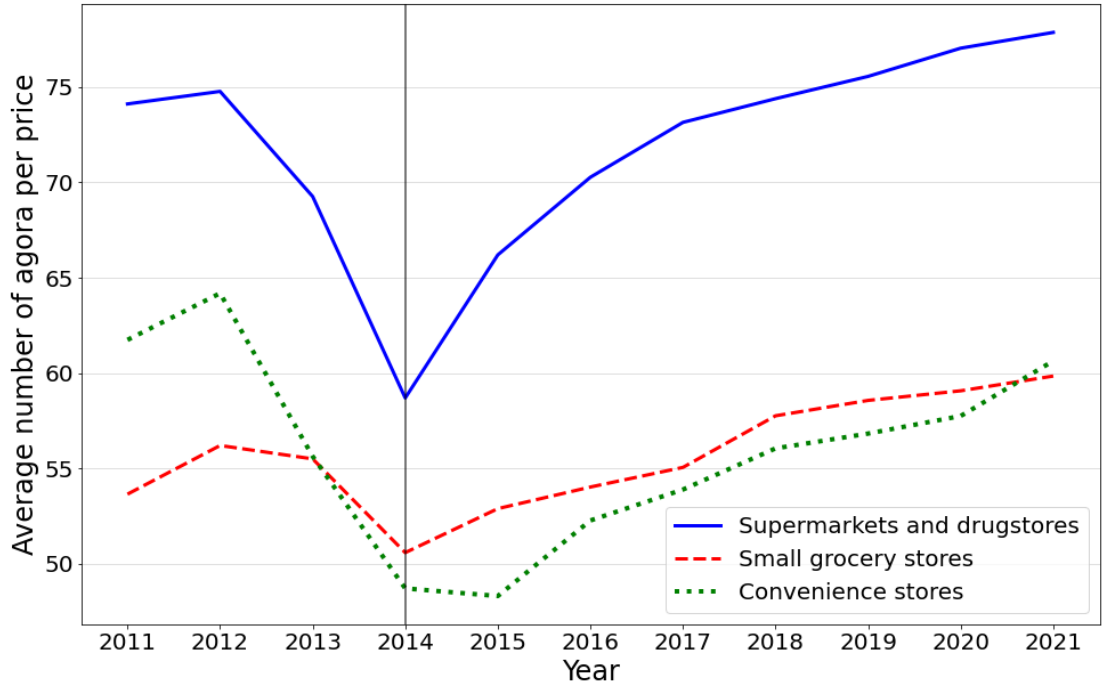

Notes: The figure depicts the average number of agora (pennies) paid per price during 2011–2021. Data is based on the CPI sample. The vertical line marks January 1, 2014, when the price rounding regulation came into effect. The figure is drawn for prices less than NIS 20.

## Table 1. Estimation of the rounding tax

| | (1) Average rounding tax per transaction (NIS) | (2) Revenue (%) | (3) Transactions (thousands) | (4) Rounding tax (NIS): Equal shares | (5) Maximum rounding tax | (6) Minimum rounding tax |
|---|---|---|---|---|---|---|
| Supermarkets and drugstores | 0.0075 | 83.80 | 188,856 | 353,962 | 150,371 | 422,390 |
| Small groceries | 0.0058 | 15.30 | 98,822 | 143,836 | 575,343 | 0.0 |
| Convenience stores | 0.0048 | 0.80 | 7,856 | 9,482 | 37,926 | 0.0 |
| Total | | 100.00 | 295,533 | 507,280 | 763,641 | 422,390 |

Notes: The table shows the calculation of the rounding tax paid by consumers before 2014. Column 1 gives the average amount paid by consumers per transaction, based on our simulation. Column 2 gives the share of each type of store in total FMCG revenue. Column 3 gives the annual number of FMCG transactions per store type. Column 4 gives the total annual rounding tax paid by shoppers under the assumption that in all stores, cash transactions comprise 25% of all transactions. Columns 5 (6) gives the maximum (minimum) possible rounding tax when we keep the total share of cash transactions at 25% but allow its share to vary across store types. The maximum rounding tax is obtained when the share of cash transactions in superstores and drugstores is 10.6%, and 100% in small grocery stores and convenience stores. The minimum rounding tax is obtained when the share of cash transactions is 29.8% in supermarkets and drugstores, and 0% in small grocery stores and convenience stores.

## Table 2. Estimation of the left-digit bias

|  | (1) | (2) | (3) | (4) |
|---|---|---|---|---|
| $\hat{\beta}^{90}$ | 0.031 | 0.038 | 0.068 | 0.081 |
| $\hat{\beta}^{00}$ | 0.020 | 0.011 | 0.031 | 0.075 |
| $\hat{\varepsilon}$ | −0.650 | −0.670 | −0.680 | −0.700 |
| $\bar{p}$ | 12.700 | 7.560 | 12.610 | 12.750 |
| $\hat{\theta}$ | 0.220 | 0.300 | 0.690 | 0.110 |
| No. of observations | 994,459 | 260,888 | 6,947,640 | 908,306 |

<u>Notes</u>: Results of OLS regressions. The dependent variable in all regressions is the log of the quantity purchased. $\hat{\beta}^{90}$ is the estimated coefficient of a dummy variable $D^{90}$ which equals 1 if the price in 2014 was NIS 0.09 lower than the 2013 modal 99-ending price. $\hat{\beta}^{00}$ is the estimated coefficient of a dummy variable $D^{00}$ which equals 1 if the price in 2014 was NIS 0.01 higher than the 2013 modal 99-ending price. $\hat{\varepsilon}$ is the estimated coefficient of the log of the price. $\bar{p}$ is the average price. $\hat{\theta}$ is the estimated coefficient of the left-digit bias, calculated as $\hat{\theta} \approx \frac{(\hat{\beta}^{90} - \hat{\beta}^{00})}{-\hat{\varepsilon}} \bar{p}$. In column 1, we use observations on products that satisfy the following criteria: (a) their price was 99-ending at least once in 2013, (b) there is at least one observation for which $D^{90} = 1$, (c) there is at least one observation for which $D^{00} = 1$, and (d) their price is below NIS 20. In column 2, we limit the sample to products whose price is below NIS 10. In column 3, we replace restrictions (b) and (c) with the restriction that there is either at least one observation for which $D^{90} = 1$, or at least one observation for which $D^{00} = 1$. In column 4, we use only data collected before January 2020. All regressions also include a dummy for prices that in 2013 were 99-ending, and fixed effects for product-stores, for year-categories, for month-categories, and for supermarket chains.

Table 3. The effect of the increase in the average number of pennies per price on the total amount paid by shoppers, 2021 vs. 2013

|  | Supermarkets and drugstores | Small grocery stores | Convenience stores |
|---|---|---|---|
| The average number of agora per price, 2021 | 77.9 | 59.8 | 49.5 |
| The average number of agora per price, 2013 | 69.3 | 55.5 | 46.2 |
| Difference (agora) | 8.6 | 4.3 | 3.3 |
| Sales volume, units (thousands) | 2,685,251.3 | 845,403.2 | 57,628.3 |
| Total effect of the increase in the average number of pennies per price (NIS) | 230,931,612 | 36,352,338 | 1,901,734 |

Notes: The sales volume is calculated from Nielsen's dataset. The average number of agora (pennies) per price is calculated using the CPI sample. The sales volume data comes from the 2013 Nielsen data. The figures in the table are calculated for prices less than NIS 20.

**Online Supplementary Web Appendix**

# Price Setting Rules, Rounding Tax, and Inattention Penalty*

Doron Sayag
Department of Economics, Bar-Ilan University
Ramat-Gan 5290002, Israel
Doronsayag2@gmail.com

Avichai Snir
Department of Economics, Bar-Ilan University
Ramat-Gan 5290002, Israel
Snirav@biu.ac.il

Daniel Levy**
Department of Economics, Bar-Ilan University
Ramat-Gan 5290002, Israel,
Department of Economics, Emory University
Atlanta, GA 30322, USA,
ICEA, ISET at TSU, and RCEA
Daniel.Levy@biu.ac.il

Revised:
November 19, 2024

## Appendix A. Using price changes that lasted two weeks

In the paper, we calculate the left-digit bias using information on scanner-data. A possible drawback is that the prices in our data are weekly averages, which can add noise to the estimation. To minimize this concern, in the paper, we use only prices that are precise up to the penny digit. After January 1, 2014, we also require that the price is rounded up to the 10-pennies digit.

As a test of robustness, in this appendix, we also require that the price following the change lasts for at least 2 weeks. Strulov-Shlain (2023) shows that the likelihood that a "spurious" price change would last more than 1 week is mute. We then re-estimate the regressions that their results are reported in Table 2 in the paper.

Following Strulov-Shlain (2021), we take advantage of the discontinuity resulting from the price rounding regulation to estimate the left-digit bias. Consider a product whose price in 2013 was NIS 9.99, and in January 1, 2014, the price was either increased by NIS 0.01 (to NIS 10.00) or decreased by NIS 0.09 (to NIS 9.90), because the store had to comply with the price rounding regulation. Any change in the demand between these two prices (i.e., between NIS 9.90 and NIS 10.00) that is above and beyond the change implied by the price elasticity, can be attributed to a left-digit bias. We therefore estimate:

$$Ln(Q_{i,s,t}) = \alpha + \beta^{90}D_{i,s,t}^{90} + \beta^{00}D_{i,s,t}^{00} + \beta^{99}D_{i,s,t}^{99} + \varepsilon ln(P_{i,s,t}) + X'\gamma + u_{i,s,t} \qquad (A1)$$

where $Q_{i,s,t}$ is the quantity of product $i$ sold in store $s$ in week $t$, $P_{i,s,t}$ is the price of product $i$ sold in store $s$ in week $t$, and $D_{i,s,t}^{90}, D_{i,s,t}^{00}$ and $D_{i,s,t}^{99}$ are dummy variables defined as follows. First, we focus on 2013, and for each product-store pair, we find the modal 99-ending price, $Mode2013_{i,s}^{99}$. We then set $D_{i,s,t}^{99}$ equal to 1 if in 2013 $P_{i,s,t}$ was equal to $Mode2013_{i,s}^{99}$, and 0 otherwise. We set $D_{i,s,t}^{00}$ equal to 1 if in 2014 $P_{i,s,t}$ was NIS 0.01 higher than $Mode2013_{i,s}^{99}$, and 0 otherwise. We set $D_{i,s,t}^{90}$ equal to 1 if in 2014 $P_{i,s,t}$ was NIS 0.09 lower than $Mode2013_{i,s}^{99}$, and 0 otherwise. For example, if the modal 99-ending price of a product-store in 2013 was NIS 9.99, then $D_{i,s,t}^{99}$ equals 1 in each week of 2013 in which the price was NIS 9.99, $D_{i,s,t}^{00}$ equals 1 in each week of 2014 in which the price was NIS 10.00, and $D_{i,s,t}^{90}$ equals 1 in each week of 2014 in which the price was NIS 9.90. The vector of control variables $X$ includes fixed effects for product-stores, for product categories × years, for product categories × months, and for supermarket chains.

The retailer dataset is not ideal for this type of estimation, because the price data is based on average prices. To overcome this limitation, we exclude observations if the average price has more than 2 digits after the decimal point or if the price does not last for at least two weeks.[1] To demonstrate robustness, in the appendix, we present results when we include only observations where the price remained unchanged for at least two consecutive weeks because it is unlikely that a spurious price would remain unchanged for more than 1 week (Strulov-Shlain, 2023).

Strulov-Shlain (2023) limits his estimation to prices below US$12, arguing that higher prices are "big-ticket items." For our main sample, we use prices up to NIS 20. Products in this price range comprise 72% of the prices in the retailer dataset and are responsible for 88% of the revenues.[2]

Table A1 contains the estimation results of equation (A1). Column 1 reports the results when we include only products that satisfy the following three criteria: (1) their price was set at a 99-ending in 2013, (2) in 2014, at least once, their price was NIS 0.01 higher than the 2013 modal 99-ending price, and (3) in 2014, at least once, their price was NIS 0.09 lower than the 2013 modal 99-ending price. Thus, we restrict the sample to product-store pairs, where the price endings fluctuated in 2013 and 2014 between 99, 90, and 00.

We find that as predicted, the coefficient of $D_{i,s,t}^{90}$ is higher than the coefficient of $D_{i,s,t}^{00}$, suggesting that 90-ending prices lead to higher demand than the nearby 00-ending prices, and the difference is larger than predicted by the demand elasticity. We follow Strulov-Shlain (2021) to estimate the left-digit bias parameter as $\hat{\theta} \approx \frac{(\hat{\beta}^{90} - \hat{\beta}^{00})}{-\hat{\varepsilon}} \bar{p}$, where $\bar{p}$ is the average price. We find that in our data, the left-digit bias parameter is 0.21, very close to what Strulov-Shlain (2023) reports for the US.

In column 2, we restrict the sample to products with prices less than NIS 10, because for them the left-digit bias is likely to be the strongest. We find that the left-digit bias parameter equals 0.37, which is indeed higher than what we obtained for the sample

---

[1] After 2014, we exclude observations also if the two rightmost digits are not a multiple of NIS 0.10.
[2] This is a less restrictive sample than Strulov-Shlain (2021) uses, because the average price in his sample is NIS 11, compared to NIS 12.70 in our sample.

where the maximum price was NIS 20.

In column 3, we again restrict the sample to products with prices less than NIS 20. We also eliminate restrictions (2) and (3), and instead include only products whose price in 2014 was at least once NIS 0.01 higher than the 2013 modal 99-ending price, or whose price in 2014 was at least once NIS 0.09 lower than the 2013 modal 99-ending price. This increases the sample considerably, from 664,856 observations to 4,788,774 observations. However, the estimation introduces noise, because we compare the effect of price changes across products rather than within products. In this sample, we find that $\hat{\theta} \approx$ 0.60, which seems too high.

Table A1. Calculating the left-digit bias using only price changes that last at least 2 weeks

|  | (1) | (2) | (3) |
|---|---|---|---|
| $\hat{\beta}^{90}$ | 0.086 | 0.070 | 0.11 |
| $\hat{\beta}^{00}$ | 0.072 | 0.0578 | 0.072 |
| $\hat{\varepsilon}$ | $-0.86$ | $-1.04$ | $-0.79$ |
| $\bar{p}$ | 12.73 | 7.53 | 12.63 |
| $\hat{\theta}$ | 0.21 | 0.37 | 0.60 |
| Observations | 664,856 | 293,417 | 4,788,774 |

<u>Notes</u>: Results of OLS regressions. The dependent variable in all regressions is the log of the quantity purchased. $\hat{\beta}^{90}$ is the estimated coefficient of a dummy variable $D^{90}$ which equals 1 if the price in 2014 was NIS 0.09 lower than the 2013 modal 99-ending price. $\hat{\beta}^{00}$ is the estimated coefficient of a dummy variable $D^{00}$ which equals 1 if the price in 2014 was NIS 0.01 higher than the 2013 modal 99-ending price. $\hat{\varepsilon}$ is the estimated coefficient of the log of the price. $\bar{p}$ is the average price. $\hat{\theta}$ is the estimated coefficient of the left-digit bias, calculated as $\hat{\theta} \approx \frac{(\hat{\beta}^{90} - \hat{\beta}^{00})}{-\hat{\varepsilon}} \bar{p}$. In column 1, we use observations on products that satisfy the following criteria: (a) their price was 99-ending at least once in 2013, (b) there is at least one observation for which $D^{90} = 1$, (c) there is at least one observation for which $D^{00} = 1$, and (d) their price is below NIS 20. In column 2, we limit the sample to products whose price is below NIS 10. In column 3, we replace restrictions (b) and (c) with the restriction that there is either at least one observation for which $D^{90} = 1$, or at least one observation for which $D^{00} = 1$. All regressions also include fixed effects for product-stores, for year-categories, for month-categories, and for supermarket chains.

**Appendix B. Inflation of 90-ending prices vs. non-90-ending prices**

In the paper, we argue that the increase in the share of 90-ending prices is due to prices being "rounded up" to 90-ending prices." In other words, we argue that following January 1, 2014, a retailer that faced a choice between setting a price at NIS 5.50 or NIS 5.90, for example, was more likely to choose NIS 5.90 than prior to January 2014.

The alternative is that retailers that considered increasing a price such that the NIS digit would change, e.g., raising a price from NIS 5.50 to NIS 6.10, preferred to forego this price change. Instead, they increased the price to only NIS 5.90.

If the alternative is the reason that we observe more 90-ending prices following January 1, 2014, then we would expect that price changes to 90-endings would be smaller, on average, than price changes to other price endings. This is because according to the alternative explanation, a large share of the changes to 90-endings are due to retailers being unwilling to make larger changes. Changes to other digits, on the other hand, are not restricted in such a way.

To test this hypothesis, we use the CPI dataset. We use it to estimate the following regression:

$$\Delta p_{i,j,t} = \alpha + \beta 90\text{-}ending_{i,j,t} + \mu_i + \pi_j + year_t + month_t + \varepsilon_{i,j,t} \qquad \text{(B1)}$$

where $\Delta p_{i,j,t}$ is log price difference of product $i$ sold in store $j$ in month $t$, conditional on a price change between month $t$ and $t-1$. $90\text{-}ending$ is a dummy for 90-ending post-change prices. $\mu, \pi$, year and month are fixed effects for products, stores, years, and months.

Table B1 gives the coefficient estimate of the 90-ending dummy. The first column reports the results when the regression includes fixed effects only for years and months. We find that the coefficient of 0-ending prices is positive and statistically significant. On average, when the post-change price is 90-ending, the price change is 3% larger than when the price changes to a different ending.

In column 2, we add fixed effects for products. We still find that when the post-change price is 90-ending, the price change is 1% larger than when the price changes to a different ending. In column 3, we also add fixed effects for stores. The results remain almost unchanged relative to column 2.

Thus, price changes to 90-endings are, on average, larger than price changes to other endings. This is inconsistent with the hypothesis that retailers use 90-endings when they want to avoid larger price changes. It is consistent, however, with retailers preferring to set 90-ending prices rather than lower prices that have the same NIS digits.

Table B1. Coefficient of the 90-ending dummy

|  | (1) | (2) | (3) |
|---|---|---|---|
| 90-ending dummy | 0.03*** | 0.01*** | 0.01*** |
|  | (0.001) | (0.001) | (0.001) |

Notes: The dependent variable is the log price change between month $t$ and $t-1$. 90-ending dummy is a dummy that equals 1 if the post-change price was 90-ending and 0 otherwise. Column 1 also includes fixed effects for year and month. In column 2, we add fixed effects for products. In column 3, we add fixed effects for stores. All regressions are based on 241,736 observations. *** $p < 0.01$

**Appendix C. Alternative estimates of the inattention tax**

In the paper, we estimate the effect of the increase in the number of pennies per price on the total amount paid by shoppers, by computing the difference in the average number of agora (pennies) paid between the period before and after the price rounding regulation, multiplied by the number of products purchased. In the paper, we focus on prices up to NIS 20, and we compare prices in 2021 and 2013.

In this appendix, we assess the robustness of the results by, first, comparing 2021 with 2012, to check that the results are robust.

Table C1 gives the estimation results, using information on prices up to NIS 20. We find that if we focus on 2012, then the effect is smaller than when we focus on 2013. The total additional amount paid by shoppers due to the increase in the number of agora (pennies) paid per price is still significant, NIS 110,733,171, which is about 0.27% of the total revenue in the FMCG market.

As a second test, we compare 2021 with 2013, but this time we do not limit the sample to products with prices smaller than NIS 20. Table C2 gives the estimation results. The total additional amount paid by shoppers due to the increase in the number of agora (pennies) paid per price is NIS 306,967,148 which is about 0.75% of the total revenue in the FMCG market.

Table C1. The effect of the increase in the average number of pennies per price on the total amount paid by shoppers, 2021 vs. 2012

|  | Supermarkets and drugstores | Small grocery stores | Convenience stores |
|---|---|---|---|
| The average number of agora per price, 2021 | 77.9 | 59.8 | 49.5 |
| The average number of agora per price, 2012 | 74.8 | 56.2 | 56.0 |
| Difference (agora) | 3.3 | 3.6 | −6.5 |
| Sales volume, units (thousands) | 2,685,251 | 845,403 | 57,628 |
| **Total effect of the increase in the average number of pennies per price (NIS)** | **84,114,987** | **30,370,882** | **−3,752,697** |

<u>Note</u>: The sales volume is calculated from Nielsen's dataset. The average number of agora (pennies) per price is calculated using the CPI sample. The table is calculated for prices less than NIS 20.

Table C2. The effect of the increase in the average number of pennies per price on the total amount paid by shoppers, 2021 vs. 2013

|  | Supermarkets and drugstores | Small grocery stores | Convenience stores |
|---|---|---|---|
| The average number of agora per price, 2021 | 73.5 | 58.3 | 52.4 |
| The average number of agora per price, 2013 | 63 | 55.6 | 48.6 |
| Difference (agora) | 10.5 | 2.7 | 3.8 |
| Sales volume, units (thousands) | 2,685,251 | 845,403 | 57,628 |
| **Total effect of the increase in the average number of pennies per price (NIS)** | **281,951,387** | **22,825,886** | **2,189,875** |

<u>Note</u>: The sales volume is calculated from Nielsen's dataset. The average number of agora (pennies) per price is calculated using the CPI sample.